\documentclass[preprint2,twoside]{hwo}

\usepackage{graphicx}
\usepackage{textcomp}
\usepackage[table, svgnames, dvipsnames]{xcolor}

\input{hwo.h}

\begin{document}

\title{\textbf{\LARGE New Frontiers in the Study of Magnetic Massive Stars with the Habitable Worlds Observatory}}
\author {\textbf{\large Alexandre David-Uraz,$^{1}$ V\'{e}ronique Petit,$^2$ Coralie Neiner,$^3$ Jean-Claude Bouret,$^4$ Ya\"{e}l Naz\'{e},$^5$ Christiana Erba,$^6$ Miriam Garcia,$^{7}$ Kenneth Gayley,$^8$ Richard Ignace,$^9$ Ji\v{r}i Krti\v{c}ka,$^{10}$ Hugues Sana,$^{11}$ Nicole St-Louis,$^{12}$ Asif ud-Doula$^{13}$}}
\affil{$^1$\small\it Central Michigan University, Mount Pleasant, Michigan, USA}
\affil{$^2$\small\it University of Delaware, Newark, Delaware, USA}
\affil{$^3$\small\it LESIA, Paris Observatory, PSL University, CNRS, Sorbonne University, Paris-Cité University, Meudon, France}
\affil{$^4$\small\it Aix-Marseille University, CNRS, CNES, LAM, Marseille, France}
\affil{$^5$\small\it Groupe d’Astrophysique des Hautes Energies, STAR, Universit\'{e} de Li\`{e}ge, Li\`{e}ge, Belgium}
\affil{$^6$\small\it Space Telescope Science Institute, Baltimore, Maryland, USA}
\affil{$^7$\small\it Centro de Astrobiolog\'{i}a, CSIC-INTA, Madrid, Spain}
\affil{$^8$\small\it University of Iowa, Iowa City, Iowa, USA}
\affil{$^9$\small\it East Tennessee State University, Johnson City, Tennessee, USA}
\affil{$^{10}$\small\it Masaryk University, Brno, Czech Republic}
\affil{$^{11}$\small\it KU Leuven, Leuven, Belgium}
\affil{$^{12}$\small\it Universit\'{e} de Montr\'{e}al, Montr\'{e}al, Qu\'{e}bec, Canada}
\affil{$^{13}$\small\it Penn State Scranton, Scranton, Pennsylvania, USA}

\author{\footnotesize{\bf Endorsed by:}
James Barron (Queen's University at Kingston), Luca Fossati (Space Research Institute, Austrian Academy of Sciences), Alex Fullerton (Space Telescope Science Institute), Fr\'{e}d\'{e}ric Galliano (CEA Paris-Saclay), Diego Godoy-Rivera (Instituto de Astrof\'{i}sica de Canarias), Joris Josiek (ZAH/ARI, Universit\"{a}t Heidelberg), Iva Krti\v{c}kov\'{a} (Masaryk University), Ignacio Mendigut\'{i}a (Centro de Astrobiolog\'{i}a, CSIC-INTA), Javier Pascual-Granado (Institute of Astrophysics of Andalusia, CSIC), Sergio Sim\'{o}n-D\'{i}az (Instituto de Astrof\'{i}sica de Canarias), Frank Soboczenski (University of York \& King's College London), R\'{o}bert Szab\'{o} (HUN-REN CSFK Konkoly Observatory), Siyao Xu (University of Florida)
}

% This section is for ADS Processing.  There must be one line per author. Leave them commented out for the present. They will be included later.
%\paperauthor{Sample~Author1}{Author1Email@email.edu}{ORCID_Or_Blank}{Author1 Institution}{Author1 Department}{City}{State/Province}{Postal Code}{Country}
%\paperauthor{Sample~Author2}{Author2Email@email.edu}{ORCID_Or_Blank}{Author2 Institution}{Author2 Department}{City}{State/Province}{Postal Code}{Country}
%\paperauthor{Sample~Author3}{Author3Email@email.edu}{ORCID_Or_Blank}{Author3 Institution}{Author3 Department}{City}{State/Province}{Postal Code}{Country}

% Please provide entries for the Author index
%\aindex{Author, F.}
%\aindex{Author, S.}
%\aindex{Author, T.}

\begin{abstract}
  High-mass stars are notable for several reasons: they are characterized by strong \textit{winds}, which inject momentum and enriched material into their surroundings, and die spectacularly as supernovae, leaving behind compact remnants and heavy elements (such as those that make life on Earth possible). Despite their relative rarity, they play a disproportionate role in the evolution of the galaxies that host them, and likely also played a significant role in the early days of the Universe. A subset ($\sim$10\%) of these stars was also found to host magnetic fields on their surface. These fields impact their evolution, and may lead to exotic physics (e.g., heavy stellar-mass black holes, pair-instability supernovae, magnetars, etc.). However, the detection and measurement of magnetic fields is limited, due to current instrumentation, to nearby massive stars in the Milky Way. To truly understand how magnetism arises in massive stars, and what role it might have played in earlier stages of our Universe, we require next-generation hardware, such as the proposed near-infrared-to-ultraviolet spectropolarimeter Pollux, on the Habitable Worlds Observatory (HWO). In this contribution, we detail how Pollux @ HWO will enable new frontiers in the study of magnetic massive stars, delivering results that will profoundly impact the fields of stellar formation, stellar evolution, compact objects, and stellar feedback.
  \\
  \\
\end{abstract}

\vspace{2cm}

\section{Science goals}

\noindent{\textbf{Science goals: Determine the origin of magnetism in massive stars (Goal 1) and the mutual roles these stars and the magnetic fields that they host play into each other’s evolution (Goal 2).}}\\

Massive stars (stars with masses at least 8 times larger than that of the Sun) are crucial objects that bridge the gap between cosmic scales. While they sit atop the stellar scale, they impact the galaxies that host them by injecting kinetic energy and chemical enrichment to their environment via their strong, radiatively-driven winds, and upon their deaths as supernovae. They also disproportionately dominate the light budget of young galaxies (vastly outshining their much more numerous lower-mass counterparts) and are responsible for most of the ionizing flux. Finally, beyond their lives as stars, they leave behind compact remnants (black holes and neutron stars), whose study has now extended into the multi-messenger domain with the advent of gravitational wave detections (e.g., \citealt{2016PhRvL.116f1102A}). Additionally, the first generations of massive stars may have played a pivotal role in the reionization of the early Universe, and in the assembly of the first galaxies. As such, understanding the lifecycle of massive stars lies at the core of a wide variety of topics in astrophysics.
While magnetism is understood to be ubiquitous at the surface of solar-type and low-mass stars due to the strong envelope convection and differential rotation that characterize them, massive stars on the other hand were not initially expected to host surface magnetic fields given their different overall internal structure: a convective core surrounded by a radiative envelope. While their core might be able to generate strong magnetism (and does; \citealt{2016ApJ...829...92A,2022MNRAS.512L..16L}), the advection time scales for these fields to reach the surface of the stars exceed their lifetimes (e.g., \citealt{1978A&A....68...57S}). Yet, surveys carried out using ground-based spectropolarimetric facilities (e.g. MiMeS and BOB; \citealt{2016MNRAS.456....2W,2015A&A...582A..45F}) have conclusively demonstrated the existence of a population of high-mass stars with strong, detectable surface magnetic fields (via the detection of the Zeeman effect in the circular polarization of spectral lines).

The results of these surveys can be described as follows: roughly 10\% of massive stars exhibit strong (of order ~kG), simple (essentially dipolar poloidal) fields \citep{2017MNRAS.465.2432G,2019MNRAS.490..274S,2019MNRAS.483.3127S}. These fields are found to be stable over time scales of years to decades (e.g., \citealt{2014MNRAS.440..182S}), and are not generally aligned with the rotational axis of their stars. This leads to periodic variability across the electromagnetic spectrum (e.g., \citealt{2007MNRAS.375..145N,2012MNRAS.419..959O}), as a magnetic field breaks the spherical symmetry both at the surface and in the circumstellar environment of a massive star. Indeed, since high-mass stars are also characterized by strong ionized winds, these outflows get channeled by the magnetic field, leading to wind confinement and forming a circumstellar structure called a magnetosphere (effectively reducing mass-loss rates; \citealt{2002ApJ...576..413U}) and very efficient shedding of angular momentum (effectively spinning down the stellar surface, leading to extremely slow rotation in some cases; \citealt{2009MNRAS.392.1022U}). 

The origins of these magnetic fields and the reason behind their incidence remain an open question. While they are generally understood to be fossil in nature (meaning that they were generated during an earlier phase of evolution rather than being continuously produced and maintained as in the dynamo case; e.g., \citealt{1982ARA&A..20..191B}), several scenarios have been proposed to explain their initial formation, each encountering its own challenges:

\begin{itemize}

\item As a star forms, the weak magnetic fields permeating the molecular cloud from which it is born could strengthen, relaxing into the configuration that is seen in main-sequence stars. However, an important challenge lies in the apparent lack of correlation between magnetic incidence and various cluster environments, or even more dramatically, the existence of binary/multiple systems with only one (detectably) magnetic component (e.g., NU Ori; \citealt{2019MNRAS.482.3950S}).

\item Pre-main sequence Herbig Ae/Be stars show magnetic characteristics (incidence, field geometry) that are comparable to that of main-sequence stars \citep{2013MNRAS.429.1027A,2013MNRAS.429.1001A}, suggesting a mechanism that operates very early in the evolution of these objects. It is also to be noted that during its earliest stages of evolution, a star may undergo changes in its internal structure, with some intermediate-mass T Tauri stars (pre-main sequence stars that will also eventually become stars with radiative envelopes) undergoing phases with large convective envelopes and exhibiting detectable magnetic fields \citep{2019A&A...622A..72V} –- however, this explanation might only be applicable to a narrow mass range, and \textbf{spectropolarimetric observations of pre-main sequence stars remain difficult to obtain with the required precision level}.

\item More recently, binary interactions (and especially mergers) have become a very prominent scenario of field generation due to detailed numerical work suggesting that merger products could reproduce some features of massive star magnetism (e.g., \citealt{2019Natur.574..211S,2020MNRAS.495.2796S}), and more recently magnetic fields have been found in objects that have been proposed to be merger products \citep{2023Sci...381..761S,2024Sci...384..214F}. This would appear compatible with the observation that the incidence of magnetic hot stars in close binaries is low (of order 2\%; \citealt{2015IAUS..307..330A}). However, some objects defy this scenario, such as the doubly-magnetic close binary star $\epsilon$ Lupi \citep{2015MNRAS.454L...1S}, for which a double merger would be difficult to reconcile with its evolutionary stage and orbital parameters. \textbf{Further detections of magnetic fields in binary systems, and in particular in a variety of environments, would help distinguish between these various scenarios.}

\end{itemize}

It should be noted however that none of these scenarios need to account for all magnetic massive stars, and that a scenario involving multiple “channels” is a perfectly viable solution. \textbf{That said, confirming and evaluating the contribution of each channel remains a priority in order to further our understanding of massive star formation and evolution.} Furthermore, while these scenarios can lead to the initial formation of a magnetic field, the observed incidence of magnetism in high-mass stars might yet depend on another crucial factor. Indeed, while most of the fields detected on main-sequence massive stars are quite strong as mentioned above, there appears to be a lower limit on their strength on the order of 100 G. Additionally, a few A stars have been discovered to harbor ultra-weak fields (of order $\sim$1 G and below), hinting at what has been termed a “magnetic desert” (or in other words a range of field strength values for which no field is observed; \citealt{2007A&A...475.1053A}). This has been interpreted as a result of a bimodality: fields over a certain critical threshold strength can remain stable over evolutionary time scales, while weaker fields are destroyed (probably by a thin, sub-surface convection zone – the so-called “FeCZ”) and replaced with ultra-weak, dynamo-generated fields \citep{2021ApJ...923..104J}. \textbf{While such fields have yet to be observed in more massive O- and B-type stars, this is possibly a result of their broader lines rendering magnetic detection more difficult, thus outlining the need for ultra-deep magnetometric observations of seemingly “non-magnetic” stars in this mass regime.} Additionally, since the sub-surface convection zone is driven by an opacity bump in iron-group elements, metallicity may play a crucial role in setting the overall characteristics of magnetism in massive stars (e.g., incidence, field strength distribution/thresholds); however, this cannot currently be ascertained given that our current instrumental capabilities do not allow us to place useful constraints on magnetic field strengths for stars in the Magellanic Clouds \citep{2020A&A...635A.163B}, i.e. at metallicities equivalent to those at the peak of star-formation in the Universe. \textbf{Therefore, an exploration of massive star magnetism in environments with different metallicities represents an exciting and necessary new frontier for the study of its origin.}

Finally, for the detected strong magnetic fields with obvious dynamical effects, the co-evolution of the field and its host star has yet to be fully constrained. Current evolutionary models of magnetic massive stars (e.g., \citealt{2019MNRAS.485.5843K,2020MNRAS.493..518K,2022MNRAS.517.2028K}), while highlighting the combined effects of magnetism on properties such as mass loss, rotation, and chemical mixing, currently assume magnetic flux conservation throughout the entire main sequence. Testing this assumption requires \textbf{deep spectropolarimetric observations of stars at various evolutionary stages}, as even with flux conservation the surface field strength is expected to decrease by an order of magnitude over the course of a star’s main sequence lifetime, and even more in later stages (e.g., \citealt{2018MNRAS.475.1521M}). Testing the models themselves requires \textbf{careful correlations of key parameters (mass loss properties, rotation rates, surface abundances) with both ages and magnetic properties}.

\textit{All of the science goals mentioned above will be uniquely enabled by Pollux, thanks to its space-borne simultaneous ultraviolet to near-infrared high-resolution spectropolarimetric capabilities. Combined with the immense light-gathering power of HWO, Pollux will deliver unprecedented views of the magnetic fields of a large sample of hot stars, and for the first time over a range of metallicities, both at the stellar photosphere and extending into the circumstellar material (via the ultraviolet). \textbf{Pollux is therefore a novel and unique opportunity to investigate the full dynamic nature of stellar magnetism in massive stars.}}\\

\noindent \underline{Overarching science questions:}

\begin{itemize}

\item What are the mechanisms that produce magnetic fields in massive stars?
\item Why are only 10\% of hot, massive stars in our Galaxy observed to have strong, stable surface magnetic fields?
\item How does (subsolar) metallicity affect the characteristics (including the incidence) of magnetic stars?
\item Do (seemingly) non-magnetic stars host small-scale weak magnetic fields, and what would such fields tell us about the internal structure of massive stars?
\item What is the interplay between binarity and magnetism, and why does the incidence of strong surface magnetic fields appear to be lower among stars in multiple systems?
\item How do these magnetic fields modify stellar evolution? 
\item How do these magnetic fields modify massive star feedback?
\item How do these fields evolve themselves (e.g. flux conservation vs. decay, nature of magnetic fields in evolved stars)?

\end{itemize}

\noindent \underline{Related ‘big board’ Astro 2020 topics:}\\

\noindent Priority Science Area: New windows on the dynamic Universe

\begin{itemize}

\item Understanding the progenitors of neutron stars and black holes

\end{itemize}

\noindent Priority Science Area: Unveiling the drivers of Galaxy growth

\begin{itemize}

\item Constraining feedback processes for various populations of massive stars

\end{itemize}

\noindent \underline{Critical for other fundamental science cases:}

\begin{itemize}

\item Current understanding of stellar formation is incomplete if it cannot account for $\sim$10\% of strongly magnetic massive stars with large-scale fields in the Galaxy.
\item Magnetic fields modify the interaction of stars with their circumstellar environment, also modifying the transfer of flux, momentum, and chemically enriched material to the environment surrounding magnetic massive stars.
\item The effect of magnetic fields on massive star evolution can propagate to the properties of their compact remnants. Indeed, magnetic massive stars are proposed to be the progenitors of pair-instability supernovae \citep{2017A&A...599L...5G}, heavy stellar-mass black holes at solar metallicities \citep{2017MNRAS.466.1052P}, and magnetars (and by extension Gamma-Ray Bursts).
\item A better picture of magnetism in massive stars at low metallicities will help understand the role that magnetism might have played in the early Universe, within the first generations of stars.
\item (Surface) magnetic fields can act as a probe of internal stellar structure (e.g., \citealt{2013MNRAS.433.2497S}).

\end{itemize}

\section{Science objectives}

\noindent{\textbf{Objective 1 (Goal 1) - Obtain \textit{extremely} deep spectropolarimetric observations of statistically representative samples of Galactic main-sequence stars (and their ``progenitors'').}}\\

\noindent \underline{Observational objective:} To observe 30 “non-magnetic” stars in the Milky Way with a magnetic sensitivity of 0.1 G. To observe 30 “non-magnetic” PMS stars in the Milky Way with a magnetic sensitivity of 1 G. To observe 30 suspected post-merger objects in the Milky Way with a sensitivity of 1 G. Each of these stars should ideally be observed 2-3 times, to mitigate averaging effects or unfavorable magnetic phases, while also exploring variability.\\

\begin{figure*}[ht]
\begin{center}
\includegraphics[width=0.8\textwidth]{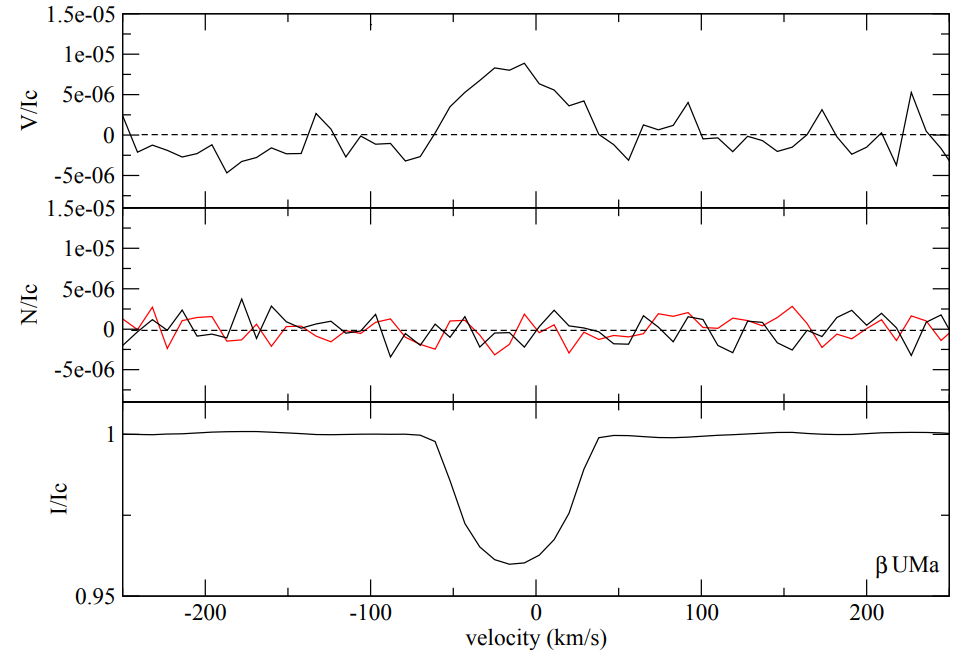}
\caption{\small Excerpt from Fig.~1 in \citet{2016A&A...586A..97B}, showing Least-Squares Deconvolution line profiles (this method is further described in Step 3) in circular polarization (top) and integrated light (bottom), as well as two diagnostic null profiles (used to detect any spurious signatures and characterize the noise level in the circularly polarized profile). Note the very high polarimetric precision required to achieve a detection, in this case in the A1 star $\beta$ UMa ($|\langle B_\textrm{z}\rangle| \sim$ 1 G).
\label{fig1}
}
\end{center}
\end{figure*}

\noindent \underline{How it addresses the science goal:} The observations will follow the procedure of, e.g., \citet{2016A&A...586A..97B} (see Fig.~\ref{fig1}), targeting bright and slow-rotating stars with stellar parameters that cover O and B main-sequence stars. If these ultra-weak fields are ubiquitous, a sample of 30 stars will enable us to differentiate it from the 10\% incidence of fossil fields. This sample of stars will allow us to critically assess the possibility of a bimodality mechanism to explain the magnetic desert.  The correlation between the field characteristics and the stellar parameters will allow us to evaluate the role of the FeCZ in the generation of these fields, and to assess scenarios for initial field generation. To push the detection limits even further, we ideally would want to obtain 2-3 observations per star. Following the Bayesian methodology of \citet{2012MNRAS.420..773P}, we can significantly improve upper limits on dipolar field strengths by obtaining a few observations at random samples (as well as improve detection probability should the field be detectable). Obtaining very deep spectropolarimetric observations of candidate magnetic star progenitors (pre-main-sequence stars and suspected post-merger objects) will also allow us to assess each of these potential field formation scenarios while also evaluating the assumed role of the aforementioned bimodality mechanism.\\

\noindent{\textbf{Objective 2 (Goal 1) - Obtain snapshot spectropolarimetric observations of a statistically representative sample of massive stars at low metallicities, both seemingly single and binary/multiple.}}\\

\noindent \underline{Observational objective:} To observe hundreds of extra-galactic stars (in the Magellanic Clouds and beyond) with a longitudinal magnetic field sensitivity of at least 100 G. Each of these stars should ideally be observed 2-3 times.\\

\begin{figure*}[ht]
\begin{center}
\includegraphics[width=0.8\textwidth]{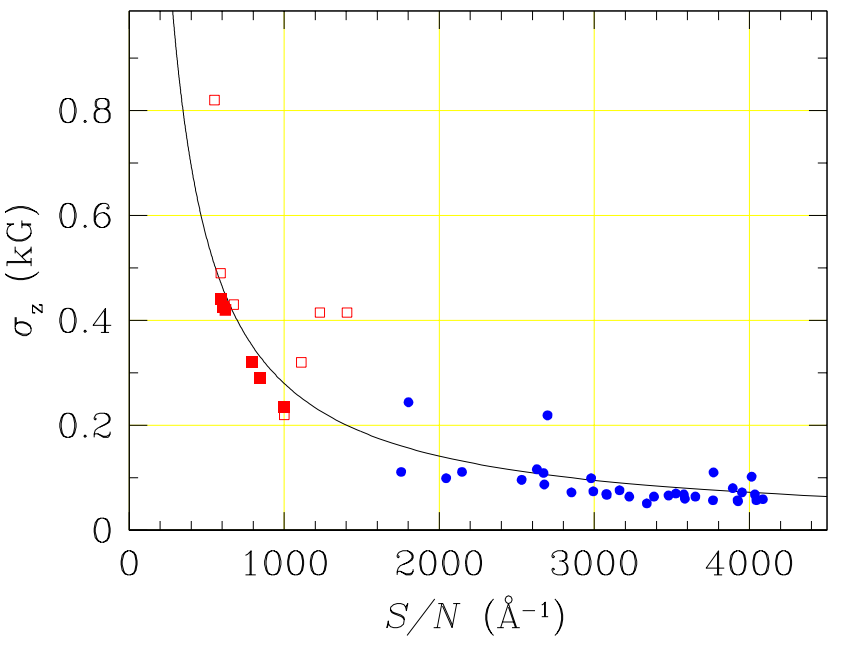}
\caption{\small Excerpt from Fig.~4 in \citet{2020A&A...635A.163B}, showing magnetic sensitivities (on the vertical axis) obtained using the FORS1/2 low-resolution spectropolarimetric instruments at ESO, for both Galactic O-type stars (blue circles), and for O-type stars in the Magellanic Clouds (red filled and empty squares). We see that for Galactic stars, typical sensitivities of order 100 G are obtained, and have been sufficient to achieve some magnetic detections. No such detection has yet been achieved for a star in the Magellanic Clouds, thus justifying our more stringent (minimal) observational constraint.
\label{fig2}
}
\end{center}
\end{figure*}

\noindent \underline{How it addresses the science goal:} The magnetic characteristics of massive stars outside of our galaxy are currently unknown. This is because, to date, no magnetic field has been observed on an extragalactic massive star (\citealt{2020A&A...635A.163B}; see Fig.~\ref{fig2}), although there is good reason to believe that such fields exist (most notably due to the spectroscopic detection of Of?p stars, a spectral subclass whose Galactic members have been consistently found to host strong magnetic fields; \citealt{2017MNRAS.465.2432G}). This objective will use, as a starting point, the sample assembled for HST’s Ultraviolet Legacy Library of Young Stars as Essential Standards (ULLYSES) program (which contained over 150 early O to B1 dwarfs and giants, as well as early O to B9 supergiants, with at least four stars per bin of spectral type and luminosity class), but will expand on that sample as described later on. The space-based spectropolarimetric capabilities of Pollux will allow us to put constraints on the incidence and characteristics of fossil fields in massive stars at low metallicity, and to evaluate the role of the FeCZ on, e.g., the magnetic strength threshold for the magnetic desert. Roughly a hundred targets would help us reliably recover a similar incidence of magnetism as in the Milky Way; however, by observing a few hundred stars we should be able to recover the magnetic incidence even if it were to be lower at lower metallicities.\\

\noindent{\textbf{Objective 3 (Goal 2) - Obtain \textit{extremely} deep spectropolarimetric observations of evolved stars and quantify the interaction of fields and other stellar properties (e.g. winds).}}\\

\noindent \underline{Observational objective:} To observe 100 evolved massive stars with a magnetic sensitivity of 0.1 G. Each of these stars should ideally be observed 2-3 times.\\

\begin{figure}[ht]
\includegraphics[width=\columnwidth]{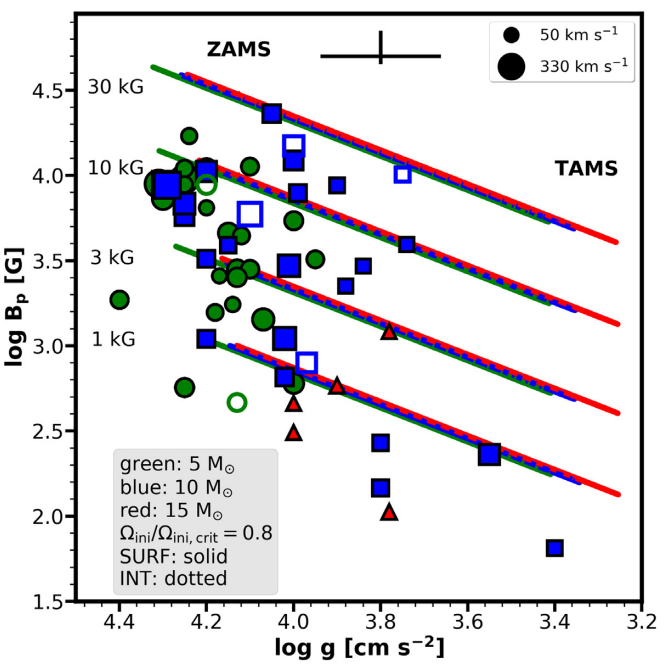}
\caption{\small Evolutionary tracks (excerpted from Fig.~5 of \citealt{2020MNRAS.493..518K}) showing a decrease in surface dipolar field strength as a star evolves across the main sequence (for various initial field strengths), assuming that the magnetic flux is conserved. Typical variations are roughly one order of magnitude, and post-main sequence stars such as the ones we aim to observe in this objective are expected to host weaker fields yet.
\label{fig3}
}
\end{figure}

\noindent \underline{How it addresses the science goal:} By determining the magnetic characteristics of evolved massive stars and comparing them to those of PMS and MS stars, we will test various scenarios of field evolution (e.g., flux conservation or decay). In particular, sufficiently detailed magnetic geometries would allow us to test whether magnetic fields become “simpler” over time, a key prediction expected if they undergo Ohmic dissipation (as the time scales associated with this mechanism are shorter for higher-order multipolar field components). The high magnetic sensitivity (and ideally repeated observations) should also allow us to assess the potential of dynamo-generated fields existing at some late evolutionary stages. Detailed analyses of the spectra of the stars will also allow us to better constrain their evolutionary phases and ages, and help reveal any discrepancy with existing (non-magnetic) evolutionary tracks that might be attributable to magnetism (and whether existing magnetic evolutionary tracks -- e.g., Fig.~\ref{fig3} can better reproduce their properties).\\

\noindent{\textbf{Objective 4 (Goal 2) - Quantify the interaction of fields and other stellar properties (winds and rotation).}}\\

\noindent \underline{Observational objective:} Perform a full magnetic/magneto- spheric characterization of 30 bright magnetic stars (previously known, or discovered in the objectives above) with dynamically-relevant winds (spanning a range of spectral types and evolutionary stages) using time-resolved (at least 10 each, preferably 20 each) spectropolarimetric observations in the ultraviolet-to-near-infrared. The required magnetic precision will depend on the (known) strength of each star’s field.\\

\noindent \underline{How it addresses the science goal:} We can obtain sophisticated tomographic mapping of structures on the surface of stars and their magnetic fields from current ground-based observations at visible and near-infrared wavelengths. However, to understand the formation and evolution of stars, we also need to explore their circumstellar environments. Pollux offers a powerful, high-resolution, full Stokes (IQUV) spectropolarimetric capability across the UV domain (100–400 nm) to uniquely trace these structures. The large aperture of HWO and high efficiency of Pollux also allows us to extend the surface mapping to more stars, allowing us to correlate field complexity with age.

\begin{figure*}[ht]
\begin{center}
\includegraphics[width=0.8\textwidth]{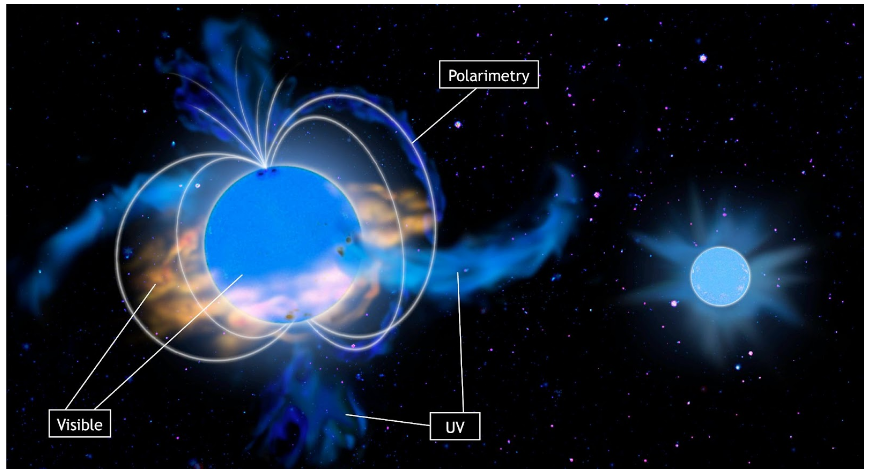}
\caption{\small Sketch of a hot star with its fossil magnetic field lines, channeled polar wind, surface spots, equatorial magnetosphere, corotating interaction regions, and a stellar companion. \textcopyright S. Cnudde.
\label{fig4}
}
\end{center}
\end{figure*}

By channeling ionised outflows (Fig.~\ref{fig4}), strong magnetic fields actually control the mass loss of massive stars, and contribute to spinning them down, as previously noted. This strongly impacts stellar structure, hence the mixing and stellar chemical yields, and ultimately the stellar evolution itself. This phenomenon has not yet been studied in detail – accurate abundances of key elements like boron, only derivable in the UV, are currently only available for a small number of stars. High-resolution UV spectropolarimetric monitoring will thus not only ensure the full mapping of the magnetosphere, but is also crucial to provide the most sensitive mass-loss and abundance diagnostics. It will lead to dramatic advances in our understanding of transport processes in (magnetic) massive stars, at all evolutionary stages (and including in multiple systems), and to a better understanding of the stellar feedback and cycle of matter.\\

\section{Physical parameters}

The physical parameters that will be used/obtained to carry out our science goals, and their current status, are presented below, and can be categorized into three main groups.

\begin{itemize}

\item Stellar parameters (luminosity, effective temperatures, surface gravity, spectroscopic masses, dynamical masses for binaries, rotation rates, mass-loss properties, abundances/metallicities, binary/multiple status): these will be used to constrain the evolutionary stage and history (e.g., merger products/binary interactions) of our targets. Given planned large ground-based and space-based spectroscopic missions that are likely to operate before the launch of HWO, it is likely that these will in fact largely be “input” parameters, which we will seek to correlate to the magnetic and magnetospheric parameters described below. That said, in some areas, Pollux should also be able to refine these determinations. In particular, refining determinations of \textbf{ages, masses, mass-loss parameters, and rotation rates} (the latter of which are directly available from the periodic variations of wind/photometric diagnostics in obliquely rotating magnetic stars; e.g. \citealt{1950MNRAS.110..395S}) could lead to important advances in our understanding of magnetic massive star evolution.

\begin{itemize}

\item Galactic (“bright”) targets: 

\begin{itemize}

\item Current state-of-the-art (across FUV-NIR range): existing HST observations + IUE catalog (for multiple observations), optical surveys (GOSSS, IACOB, SMASH, etc.)

\end{itemize}

\item (“faint”) targets in the MCs:

\begin{itemize}

\item Current state-of-the-art across FUV-NIR range: HTTP, ULLYSES in the UV + XShootU and other optical/NIR surveys; UVEX (upcoming)

\item Current state-of-the-art for multiplicity: VFTS/TMBM, BLOeM, 4MOST 1001MC (upcoming)

\end{itemize}

\end{itemize}

\item Magnetic parameters (strength, geometry): we will characterize the incidence and geometry of large-scale magnetic fields in stars in low-metallicity environments, detect ultra-weak and small-scale fields in Galactic stars (MS and evolved), and verify whether the magnetic desert threshold field strength value varies as a function of metallicity. Time series of observations will also be obtained to fully characterize the field structure in a subset of bright, favorable stars.

\begin{itemize}

\item Large-scale:

\begin{itemize}

\item Current state-of-the-art (Galactic: MiMeS/BOB; MCs: \textit{nothing!} – current best constraints from FORS2; \citealt{2020A&A...635A.163B})

\end{itemize}

\item Ultra-weak/small-scale:

\begin{itemize}

\item Current state-of-the-art: only ever observed in later-type MS stars (A-type) and evolved stars (LIFE project)

\end{itemize}

\item Detailed structure: for a subset of stars, phase-resolved spectropolarimetric datasets allow to map the surface field with high precision

\begin{itemize}

\item Current state-of-the-art: phase-resolved observations of bright stars in Stokes I/V

\end{itemize}

\end{itemize}

\item Magnetospheric parameters (density/velocity/magnetic field extrapolation): we will characterize the interaction between magnetic fields and winds in a few bright stars and test existing theoretical prescriptions for the structure of magnetospheres in various contexts.

\begin{itemize}

\item Current state-of-the-art: several phases of UV spectroscopy for some stars (HST/IUE), equivalent width curves for wind-sensitive lines (e.g., H$\alpha$) in the optical, source function field extrapolations.

\end{itemize}

\end{itemize}

\noindent \textbf{\underline{Sample definition:}}

\begin{itemize}

\item MCs (and beyond): hundreds of stars, prioritizing brightest candidates that allow us to reach the required precision for magnetic field measurements. As an initial sample, we will use the ULLYSES target list (described in more detail below), including known extragalactic Of?p stars in particular. However, it should be noted that ULLYSES is biased against binaries, so we expect to complement our target list with objects from VFTS \& BLOeM, combined with the target list for the upcoming UVEX mission.

\item Evolved stars: modeled on the LIFE target list (described in more detail below), but also extending to lower magnitudes (at least 100). We will also include the brightest and most favorable Wolf-Rayet stars – due to their optically-thick winds, field detection is much more difficult, and therefore we will aim for a more conservative magnetic sensitivity (of order a few 100 G). These are important evolutionary links to SN/GRBs, but have only been claimed to show marginal detection (with the notable exception of the extremely strongly magnetic HD 45166, understood to be a merger product; \citealt{2023Sci...381..761S}). Similarly, we will aim to include sufficiently bright targets from the expanding sample of known stripped stars (e.g. \citealt{2023Sci...382.1287D}), to further probe the effects of binary interactions.

\item “Non-magnetic” Galactic stars (PMS and MS): the brightest, most suitable (slow rotation) targets for which no magnetic detection has been achieved (despite prior spectropolarimetric observations with moderate magnetic sensitivity) will be selected.

\item Targets for detailed field/magnetosphere mapping: bright, known magnetic stars, chosen to span a range of spectral types and ages.

\end{itemize}

\noindent \textit{Notes about the ULLYSES sample:}

{\small Following the Hubble Ultraviolet Legacy Library of Young Stars as Essential Standards (ULLYSES) program, a complete reference spectral library of the largest sample of massive stars at low metallicity is now available.  More precisely, ULLYSES has obtained UV spectra of 94 targets in the Large Magellanic Cloud (LMC) and 60 stars in the Small Magellanic Cloud (SMC). The sample was designed such that it includes early O to B1 dwarfs and giants, as well as early O to B9 supergiants, with at least four stars per bin of spectral type and luminosity class. Seven WRs in the LMC and 4 in the SMC have also been observed. Finally, a handful of massive stars in galaxies beyond the Magellanic clouds (more distant but expectedly lower metallicity) have been observed, namely in Sextans-A and NGC 3109.

The ULLYSES observations complement and leverage existing archival HST data. Data obtained with FUSE (Far Ultraviolet Space Explorer) are also included whenever possible. In total, the ULLYSES sample includes 182 stars in the LMC, 133 stars in the SMC, and 3 more stars in both Sext-A and NGC 3109.  
 
Spectra have been obtained with the Cosmic Origins Spectrograph (COS) or the Space Telescope Imaging Spectrograph (STIS). For LMC and SMC stars, typical SNR= 30 for COS (R=15,000) and SNR=20 for STIS (R=30,000) have been obtained. COS low resolution setting (R~3000) has been used for stars in Sext-A and NGC 3109 because of the expensive orbit cost of spectroscopic observations for these stars.  
 
In order to make the best of this extensive dataset of low-Z massive stars, analysis tools have been developed. Extended grids of models have been computed (e.g., \citealt{2024A&A...690A.318M}) to interpret the data. Early results about the wind properties and surface abundances of selected sub-samples have been presented \citep{2024A&A...688A.105H,2024MNRAS.52711422P,2024A&A...689A..31M}. A more systematic, pipeline-type, approach, starts to deliver its first results \citep{2025A&A...695A.198B} about the physical (e.g. photospheric, mass, rotation) and wind properties, which further helps better interpret the evolutionary status of the objects.  All these tools yet ignore what the impact of magnetic fields would be on the usual surface and wind properties just mentioned above.

For instance, models are lacking to relate the presence, intensity and geometry of magnetic fields to surface abundance anomalies. In particular, how to interpret surface abundance anomalies of species such as Nitrogen to the magnetic properties of a star (or to its rotational properties).}\\

\noindent \textit{Notes about the LIFE sample:}

{\small LIFE targets were only bright objects (4 $\le V \le$ 8), with spectral types O, B, and A, visible from CFHT (i.e. with declination above -50$^\textrm{o}$), that are not Wolf-Rayet stars (because those were found to be particularly challenging in previous studies), that satisfy the following criteria:

\begin{itemize}

\item Class I stars: An original sample of O, B, and A, luminosity class I stars was extracted using SIMBAD according to the criteria above. Additional suitable targets from \citet{2010MNRAS.404.1306F}, \citet{2006A&A...446..279C}, \citet{2008A&A...478..823M}, and \citet{2012A&A...543A..80F} were added. Comparison with the model evolutionary tracks of \citet{2012A&A...537A.146E} indicates that all selected stars are bona fide post-MS objects. The observable sample consists of 60 stars, including 2 LBVs.

\item Class II and III stars: An original sample of O, B, and A, luminosity class II, and hotter than B6, luminosity class III stars according to the criteria above was extracted using SIMBAD. From this list, only stars with Hipparcos parallaxes were retained. Using Johnson photometric colours and the (B-V)-$T_\textrm{eff}$ and bolometric correction calibrations H-R diagram positions were determined. Comparing these with model evolutionary tracks of \citet{2012A&A...537A.146E}, stars with positions coincident with the MS were rejected. This latter step in particular rejects unevolved stars with peculiar spectra, e.g., HgMn stars, which were found to significantly contaminate the sample of class III objects. The observable LIFE sample finally consisted of 55 stars of class II and 35 stars of class III. 55 of them have actually been observed. 

\end{itemize}

\noindent Known magnetic evolved hot stars:

\begin{itemize}

\item 2 already known previously to LIFE : $\zeta$ Ori Aa and $\varepsilon$ CMa

\item 2 discovered through the BritePol survey : $\iota$ Car and HR 3890

\item 7 discovered with LIFE : 19 Aur, HR 3042, HIP 38584, HD 167686, $\eta$ Leo, 13 Mon, d Car

\end{itemize}

\noindent All observed fields are very weak: between 0.1 G and a few G.}\\

\noindent \textbf{\underline{Parameter overview:}} Understanding most stellar parameters to be “input” parameters which will be correlated with the magnetic properties of the stars we observe, we mostly focus on the latter, as shown in Table~\ref{tab1}.

\begin{table*}[!ht]
\caption{Physical parameters}
\smallskip
\begin{center}
{\small
\begin{tabular}{|p{0.2\textwidth}|p{0.16\textwidth}|p{0.16\textwidth}|p{0.16\textwidth}|p{0.16\textwidth}|}  % l = left, c = centered
\tableline
%\noalign{\smallskip}
\rowcolor{lightgray}
Physical Parameter & State of the Art & Incremental Progress (Enhancing) & Substantial Progress (Enabling) & Major Progress (Breakthrough)\\
%\noalign{\smallskip}
\tableline
\tableline
%\noalign{\smallskip}
Metallicity & MW only (so nearly solar) & LMC (0.5 $Z_\odot$) & SMC (0.2 $Z_\odot$) & {\color{blue}$< 0.2 Z_\odot$} \\
\tableline
Magnetic sensitivity (ultra-weak/small-scale fields + post-MS) & For OB stars, typically no less than a few tens of G & 5 G & 0.5 G & {\color{blue}0.1 G} \\
\tableline
Magnetic sensitivity (for low-Z and WR stars) & Typically several 100s of G at best (and often much worse) & 500 G & 300 G & {\color{blue}100 G} \\
%\noalign{\smallskip}
\tableline % Sometimes you just need a line between table rows
%\noalign{\smallskip}
Number of observations per star (snapshot) & Typically these will not have been observed (at a given magnetic sensitivity) & \multicolumn{2}{l|}{1} & {\color{blue}2-3} \\ % Sometimes you have empty cells
%\noalign{\smallskip} % Sometimes you just need space between table rows
\tableline
Number of observations per star (phase-resolved) & For already known magnetic stars, they will have typically been observed a handful of times at most & \multicolumn{2}{l|}{10 observations well spaced in rotational phase} & {\color{blue}20 observations well spaced in rotational phase} \\% No \\ if the last row
%\noalign{\smallskip}
\tableline
\end{tabular}\label{tab1}
}
\end{center}
\end{table*}

\section{Step 4: Description of observations}

\noindent The observations will be divided into two main components, as described above:

\begin{itemize}

\item Survey component: to detect new magnetic fields with unprecedented magnetic sensitivity.

\item Targeted component: to fully characterize the magnetic fields of known magnetic stars and map their circumstellar environment. 

\end{itemize}

Observations will consist of FUV-NIR spectroscopy/ spectropolarimetry in full Stokes parameters (IQUV).\\

\noindent Detailed requirements:

\begin{itemize}

\item Wavelength range: $\sim$100-1600 nm (the FUV is essential to diagnose wind properties using resonance lines, as well as use the Zeeman effect, as well as possibly the Hanle effect, to map out the field into the wind).

\item Full Stokes parameters: circular polarization essential to diagnose the longitudinal component of the magnetic field through the Zeeman effect; linear polarization can also provide information on the transverse component of the field, as well as allow us to observe both the \"{O}hman (constraining rotation rates) and Hanle (independent constraint on the magnetic field – especially useful as mentioned above to map out the field in the wind using the UV) effects.

\item Number of observations per target: For the survey component, at least 1 snapshot observation, but preferably 2-3. For the targeted component, at least 10 phase-resolved observations, preferably 20.

\item Polarimetric precision: to obtain the very deep observations that we need to carry out our objectives, a minimum precision of 10$^{-4}$ is required, with a preferred value of 10$^{-6}$.

\end{itemize}

\noindent \textbf{\underline{The need for HWO:}}

Ground-based optical and near-infrared spectropolarimetric observations have already provided us with a wealth of information about magnetism in the upper HR diagram, but are also associated with a certain number of limitations. The large aperture and broad spectral range of HWO, coupled with the high-resolution spectropolarimetric capabilities of Pollux, will help us reach new frontiers in the study of magnetic massive stars: 

\begin{itemize}

\item by detecting ultra-weak and small-scale fields in Milky Way stars;

\item by detecting strong, organized fields in extragalactic stars.

\end{itemize}

The science goals and objectives detailed in this document can only be achieved through this powerful combination. Therefore, HWO is crucial to the realization of this science case.\\

\noindent \textbf{\underline{Observing requirements:}} The observing requirements are summarized in Table~\ref{tab2}.\\

\begin{table*}[!ht]
\caption{Observing requirements}
\smallskip
\begin{center}
{\small
\begin{tabular}{|p{0.2\textwidth}|p{0.16\textwidth}|p{0.16\textwidth}|p{0.16\textwidth}|p{0.16\textwidth}|}  % l = left, c = centered
\tableline
%\noalign{\smallskip}
\rowcolor{lightgray}
Observation Requirement & State of the Art & Incremental Progress (Enhancing) & Substantial Progress (Enabling) & Major Progress (Breakthrough)\\
%\noalign{\smallskip}
\tableline
\tableline
%\noalign{\smallskip}
Full Stokes (IQUV) spectropolarimetry & Stokes IV for most MW stars, low-res. IV for MCs (no detection!) & Stokes IQUV for MW stars & High-res. Stokes IV for MCs & {\color{blue}High-res. Stokes IQUV for MW and MCs} \\
\tableline
Waveband range (for spectropol.) & Optical-NIR & \multicolumn{2}{l|}{NUV to NIR} & {\color{blue}FUV to NIR} \\
%\noalign{\smallskip}
\tableline % Sometimes you just need a line between table rows
%\noalign{\smallskip}
Spectral resolution & $\sim$120,000 (optical spectropol.); $\sim$40,000 for UV spectro. (STIS) & \multicolumn{3}{p{0.48\textwidth}|}{\color{blue}Adding (high-res. $>$ 60,000) UV spectropolarimetry is a big step forward (no need to go much higher in resolution than existing optical spectropolarimetric observations)} \\ % Sometimes you have empty cells
%\noalign{\smallskip} % Sometimes you just need space between table rows
\tableline
Polarimetric precision &  $\sim$10$^{-4}$ in NIR & 10$^{-4}$ across FUV to NIR & 10$^{-5}$ across FUV to NIR & {\color{blue}10$^{-6}$ across FUV to NIR} \\% No \\ if the last row
%\noalign{\smallskip}
\tableline
\end{tabular}\label{tab2}
}
\end{center}
\end{table*}

%\begin{figure}[ht]
%\includegraphics[width=\columnwidth]{author_fig1}
%\caption{\small A cool astronomy figure.
%\label{author_fig1_label}
%}
%\end{figure}

% one that spans both columns
%\begin{figure*}[ht]
%\begin{center}
%\includegraphics[width=0.6\textwidth]{author_fig1}
%\caption{\small A cool astronomy figure.
%\label{author_fig2_label}
%}
%\end{center}
%\end{figure*}

%\begin{table*}[!ht]
%\caption{Tables in \LaTeXe}
%\smallskip
%\begin{center}
%{\small
%\begin{tabular}{llc}  % l = left, c = centered
%\tableline
%\noalign{\smallskip}
%First Column & Second Column & Third Column:\\
%\noalign{\smallskip}
%\tableline
%\noalign{\smallskip}
%First Row, First Column & First Row, Second Column & First Row, Third Column \\
%Second Row, First Column & Second Row, Second Column & Second Row, Third Column \\
%Third Row, First Column & Third Row, Second Column & Third Row, Third Column \\
%\noalign{\smallskip}
%\tableline % Sometimes you just need a line between table rows
%\noalign{\smallskip}
%Fourth Row, First Column & ~ & Fourth Row, Third Column \\ % Sometimes you have empty cells
%\noalign{\smallskip} % Sometimes you just need space between table rows
%Fifth Row First Column & Fifth Row, Second Column & Fifth Row, Third Column\\ % No \\ if the last row
%\noalign{\smallskip}
%\tableline\
%\end{tabular}
%}
%\end{center}
%\end{table*}
%\noindent These tables can get a little messy, but this format is the most common.

%\acknowledgements
{\bf Acknowledgements.} CE gratefully acknowledges support for this work provided by NASA through grant numbers HST-AR-15794 and HST-GO-16767 from the Space Telescope Science Institute, which is operated by AURA, Inc., under NASA contract NAS 5-26555.  

\bibliography{David-Uraz.bib}

\begin{thebibliography}{}
\parskip=0pt \itemsep=0pt \small \baselineskip=11pt
\expandafter\ifx\csname natexlab\endcsname\relax\def\natexlab#1{#1}\fi
\providecommand{\url}[1]{\href{#1}{#1}}
\providecommand{\dodoi}[1]{}
\providecommand{\doeprint}[1]{\href{http://ascl.net/#1}{#1}}
\providecommand{\doarXiv}[1]{\href{https://arxiv.org/abs/#1}{arXiv:#1}}

\bibitem[{{Abbott} {et~al.}(2016){Abbott}, {Abbott}, {Abbott}, {Abernathy}, {Acernese}, {Ackley}, {Adams}, {Adams}, {Addesso}, {Adhikari}, {Adya}, {Affeldt}, {Agathos}, {Agatsuma}, {Aggarwal}, {Aguiar}, {Aiello}, {Ain}, {Ajith}, {Allen}, {Allocca}, {Altin}, {Anderson}, {Anderson}, {Arai}, {Arain}, {Araya}, {Arceneaux}, {Areeda}, {Arnaud}, {Arun}, {Ascenzi}, {Ashton}, {Ast}, {Aston}, {Astone}, {Aufmuth}, {Aulbert}, {Babak}, {Bacon}, {Bader}, {Baker}, {Baldaccini}, {Ballardin}, {Ballmer}, {Barayoga}, {Barclay}, {Barish}, {Barker}, {Barone}, {Barr}, {Barsotti}, {Barsuglia}, {Barta}, {Bartlett}, {Barton}, {Bartos}, {Bassiri}, {Basti}, {Batch}, {Baune}, {Bavigadda}, {Bazzan}, {Behnke}, {Bejger}, {Belczynski}, {Bell}, {Bell}, {Berger}, {Bergman}, {Bergmann}, {Berry}, {Bersanetti}, {Bertolini}, {Betzwieser}, {Bhagwat}, {Bhandare}, {Bilenko}, {Billingsley}, {Birch}, {Birney}, {Birnholtz}, {Biscans}, {Bisht}, {Bitossi}, {Biwer}, {Bizouard}, {Blackburn}, {Blair}, {Blair}, {Blair}, {Bloemen}, {Bock}, {Bodiya}, {Boer},
  {Bogaert}, {Bogan}, {Bohe}, {Bojtos}, {Bond}, {Bondu}, {Bonnand}, {Boom}, {Bork}, {Boschi}, {Bose}, {Bouffanais}, {Bozzi}, {Bradaschia}, {Brady}, {Braginsky}, {Branchesi}, {Brau}, {Briant}, {Brillet}, {Brinkmann}, {Brisson}, {Brockill}, {Brooks}, {Brown}, {Brown}, {Brown}, {Buchanan}, {Buikema}, {Bulik}, {Bulten}, {Buonanno}, {Buskulic}, {Buy}, {Byer}, {Cabero}, {Cadonati}, {Cagnoli}, {Cahillane}, {Bustillo}, {Callister}, {Calloni}, {Camp}, {Cannon}, {Cao}, {Capano}, {Capocasa}, {Carbognani}, {Caride}, {Diaz}, {Casentini}, {Caudill}, {Cavagli{\`a}}, {Cavalier}, {Cavalieri}, {Cella}, {Cepeda}, {Baiardi}, {Cerretani}, {Cesarini}, {Chakraborty}, {Chalermsongsak}, {Chamberlin}, {Chan}, {Chao}, {Charlton}, {Chassande-Mottin}, {Chen}, {Chen}, {Cheng}, {Chincarini}, {Chiummo}, {Cho}, {Cho}, {Chow}, {Christensen}, {Chu}, {Chua}, {Chung}, {Ciani}, {Clara}, {Clark}, {Cleva}, {Coccia}, {Cohadon}, {Colla}, {Collette}, {Cominsky}, {Constancio}, {Conte}, {Conti}, {Cook}, {Corbitt}, {Cornish}, {Corsi}, {Cortese}, {Costa},
  {Coughlin}, {Coughlin}, {Coulon}, {Countryman}, {Couvares}, {Cowan}, {Coward}, \& {Cowart}}]{2016PhRvL.116f1102A}
{Abbott}, B.~P., {Abbott}, R., {Abbott}, T.~D., {et~al.} 2016, \href{http://doi.org/10.1103/PhysRevLett.116.061102}{\color{blue}\prl}, \href{https://ui.adsabs.harvard.edu/abs/2016PhRvL.116f1102A}{\color{blue}116}, 061102

\bibitem[{{Alecian} {et~al.}(2013{\natexlab{a}}){Alecian}, {Wade}, {Catala}, {Grunhut}, {Landstreet}, {B{\"o}hm}, {Folsom}, \& {Marsden}}]{2013MNRAS.429.1027A}
{Alecian}, E., {Wade}, G.~A., {Catala}, C., {et~al.} 2013{\natexlab{a}}, \href{http://doi.org/10.1093/mnras/sts384}{\color{blue}\mnras}, \href{https://ui.adsabs.harvard.edu/abs/2013MNRAS.429.1027A}{\color{blue}429}, 1027

\bibitem[{{Alecian} {et~al.}(2013{\natexlab{b}}){Alecian}, {Wade}, {Catala}, {Grunhut}, {Landstreet}, {Bagnulo}, {B{\"o}hm}, {Folsom}, {Marsden}, \& {Waite}}]{2013MNRAS.429.1001A}
---. 2013{\natexlab{b}}, \href{http://doi.org/10.1093/mnras/sts383}{\color{blue}\mnras}, \href{https://ui.adsabs.harvard.edu/abs/2013MNRAS.429.1001A}{\color{blue}429}, 1001

\bibitem[{{Alecian} {et~al.}(2015){Alecian}, {Neiner}, {Wade}, {Mathis}, {Bohlender}, {C{\'e}bron}, {Folsom}, {Grunhut}, {Le Bouquin}, {Petit}, {Sana}, {Tkachenko}, \& {ud-Doula}}]{2015IAUS..307..330A}
{Alecian}, E., {Neiner}, C., {Wade}, G.~A., {et~al.} 2015, in IAU Symposium, Vol. 307, \href{http://doi.org/10.1017/S1743921314007030}{\color{blue}New Windows on Massive Stars}, ed. G.~{Meynet}, C.~{Georgy}, J.~{Groh}, \& P.~{Stee}, 330--335

\bibitem[{{Augustson} {et~al.}(2016){Augustson}, {Brun}, \& {Toomre}}]{2016ApJ...829...92A}
{Augustson}, K.~C., {Brun}, A.~S., \& {Toomre}, J. 2016, \href{http://doi.org/10.3847/0004-637X/829/2/92}{\color{blue}\apj}, \href{https://ui.adsabs.harvard.edu/abs/2016ApJ...829...92A}{\color{blue}829}, 92

\bibitem[{{Auri{\`e}re} {et~al.}(2007){Auri{\`e}re}, {Wade}, {Silvester}, {Ligni{\`e}res}, {Bagnulo}, {Bale}, {Dintrans}, {Donati}, {Folsom}, {Gruberbauer}, {Hui Bon Hoa}, {Jeffers}, {Johnson}, {Landstreet}, {L{\`e}bre}, {Lueftinger}, {Marsden}, {Mouillet}, {Naseri}, {Paletou}, {Petit}, {Power}, {Rincon}, {Strasser}, \& {Toqu{\'e}}}]{2007A&A...475.1053A}
{Auri{\`e}re}, M., {Wade}, G.~A., {Silvester}, J., {et~al.} 2007, \href{http://doi.org/10.1051/0004-6361:20078189}{\color{blue}\aap}, \href{https://ui.adsabs.harvard.edu/abs/2007A&A...475.1053A}{\color{blue}475}, 1053

\bibitem[{{Bagnulo} {et~al.}(2020){Bagnulo}, {Wade}, {Naz{\'e}}, {Grunhut}, {Shultz}, {Asher}, {Crowther}, {Evans}, {David-Uraz}, {Howarth}, {Morrell}, {Munoz}, {Neiner}, {Puls}, {Szyma{\'n}ski}, \& {Vink}}]{2020A&A...635A.163B}
{Bagnulo}, S., {Wade}, G.~A., {Naz{\'e}}, Y., {et~al.} 2020, \href{http://doi.org/10.1051/0004-6361/201937098}{\color{blue}\aap}, \href{https://ui.adsabs.harvard.edu/abs/2020A&A...635A.163B}{\color{blue}635}, A163

\bibitem[{{Bestenlehner} {et~al.}(2025){Bestenlehner}, {Crowther}, {Hawcroft}, {Sana}, {Tramper}, {Vink}, {Brands}, {Sander}, \& {XShootU Collaboration}}]{2025A&A...695A.198B}
{Bestenlehner}, J.~M., {Crowther}, P.~A., {Hawcroft}, C., {et~al.} 2025, \href{http://doi.org/10.1051/0004-6361/202452491}{\color{blue}\aap}, \href{https://ui.adsabs.harvard.edu/abs/2025A&A...695A.198B}{\color{blue}695}, A198

\bibitem[{{Blaz{\`e}re} {et~al.}(2016){Blaz{\`e}re}, {Petit}, {Ligni{\`e}res}, {Auri{\`e}re}, {Ballot}, {B{\"o}hm}, {Folsom}, {Gaurat}, {Jouve}, {Lopez Ariste}, {Neiner}, \& {Wade}}]{2016A&A...586A..97B}
{Blaz{\`e}re}, A., {Petit}, P., {Ligni{\`e}res}, F., {et~al.} 2016, \href{http://doi.org/10.1051/0004-6361/201527556}{\color{blue}\aap}, \href{https://ui.adsabs.harvard.edu/abs/2016A&A...586A..97B}{\color{blue}586}, A97

\bibitem[{{Borra} {et~al.}(1982){Borra}, {Landstreet}, \& {Mestel}}]{1982ARA&A..20..191B}
{Borra}, E.~F., {Landstreet}, J.~D., \& {Mestel}, L. 1982, \href{http://doi.org/10.1146/annurev.aa.20.090182.001203}{\color{blue}\araa}, \href{https://ui.adsabs.harvard.edu/abs/1982ARA&A..20..191B}{\color{blue}20}, 191

\bibitem[{{Crowther} {et~al.}(2006){Crowther}, {Lennon}, \& {Walborn}}]{2006A&A...446..279C}
{Crowther}, P.~A., {Lennon}, D.~J., \& {Walborn}, N.~R. 2006, \href{http://doi.org/10.1051/0004-6361:20053685}{\color{blue}\aap}, \href{https://ui.adsabs.harvard.edu/abs/2006A&A...446..279C}{\color{blue}446}, 279

\bibitem[{{Drout} {et~al.}(2023){Drout}, {G{\"o}tberg}, {Ludwig}, {Groh}, {de Mink}, {O'Grady}, \& {Smith}}]{2023Sci...382.1287D}
{Drout}, M.~R., {G{\"o}tberg}, Y., {Ludwig}, B.~A., {et~al.} 2023, \href{http://doi.org/10.1126/science.ade4970}{\color{blue}Science}, \href{https://ui.adsabs.harvard.edu/abs/2023Sci...382.1287D}{\color{blue}382}, 1287

\bibitem[{{Ekstr{\"o}m} {et~al.}(2012){Ekstr{\"o}m}, {Georgy}, {Eggenberger}, {Meynet}, {Mowlavi}, {Wyttenbach}, {Granada}, {Decressin}, {Hirschi}, {Frischknecht}, {Charbonnel}, \& {Maeder}}]{2012A&A...537A.146E}
{Ekstr{\"o}m}, S., {Georgy}, C., {Eggenberger}, P., {et~al.} 2012, \href{http://doi.org/10.1051/0004-6361/201117751}{\color{blue}\aap}, \href{https://ui.adsabs.harvard.edu/abs/2012A&A...537A.146E}{\color{blue}537}, A146

\bibitem[{{Firnstein} \& {Przybilla}(2012)}]{2012A&A...543A..80F}
{Firnstein}, M., \& {Przybilla}, N. 2012, \href{http://doi.org/10.1051/0004-6361/201219034}{\color{blue}\aap}, \href{https://ui.adsabs.harvard.edu/abs/2012A&A...543A..80F}{\color{blue}543}, A80

\bibitem[{{Fossati} {et~al.}(2015){Fossati}, {Castro}, {Sch{\"o}ller}, {Hubrig}, {Langer}, {Morel}, {Briquet}, {Herrero}, {Przybilla}, {Sana}, {Schneider}, {de Koter}, \& {BOB Collaboration}}]{2015A&A...582A..45F}
{Fossati}, L., {Castro}, N., {Sch{\"o}ller}, M., {et~al.} 2015, \href{http://doi.org/10.1051/0004-6361/201526725}{\color{blue}\aap}, \href{https://ui.adsabs.harvard.edu/abs/2015A&A...582A..45F}{\color{blue}582}, A45

\bibitem[{{Fraser} {et~al.}(2010){Fraser}, {Dufton}, {Hunter}, \& {Ryans}}]{2010MNRAS.404.1306F}
{Fraser}, M., {Dufton}, P.~L., {Hunter}, I., {et~al.} 2010, \href{http://doi.org/10.1111/j.1365-2966.2010.16392.x}{\color{blue}\mnras}, \href{https://ui.adsabs.harvard.edu/abs/2010MNRAS.404.1306F}{\color{blue}404}, 1306

\bibitem[{{Frost} {et~al.}(2024){Frost}, {Sana}, {Mahy}, {Wade}, {Barron}, {Le Bouquin}, {M{\'e}rand}, {Schneider}, {Shenar}, {Barb{\'a}}, {Bowman}, {Fabry}, {Farhang}, {Marchant}, {Morrell}, \& {Smoker}}]{2024Sci...384..214F}
{Frost}, A.~J., {Sana}, H., {Mahy}, L., {et~al.} 2024, \href{http://doi.org/10.1126/science.adg7700}{\color{blue}Science}, \href{https://ui.adsabs.harvard.edu/abs/2024Sci...384..214F}{\color{blue}384}, 214

\bibitem[{{Georgy} {et~al.}(2017){Georgy}, {Meynet}, {Ekstr{\"o}m}, {Wade}, {Petit}, {Keszthelyi}, \& {Hirschi}}]{2017A&A...599L...5G}
{Georgy}, C., {Meynet}, G., {Ekstr{\"o}m}, S., {et~al.} 2017, \href{http://doi.org/10.1051/0004-6361/201730401}{\color{blue}\aap}, \href{https://ui.adsabs.harvard.edu/abs/2017A&A...599L...5G}{\color{blue}599}, L5

\bibitem[{{Grunhut} {et~al.}(2017){Grunhut}, {Wade}, {Neiner}, {Oksala}, {Petit}, {Alecian}, {Bohlender}, {Bouret}, {Henrichs}, {Hussain}, {Kochukhov}, \& {MiMeS Collaboration}}]{2017MNRAS.465.2432G}
{Grunhut}, J.~H., {Wade}, G.~A., {Neiner}, C., {et~al.} 2017, \href{http://doi.org/10.1093/mnras/stw2743}{\color{blue}\mnras}, \href{https://ui.adsabs.harvard.edu/abs/2017MNRAS.465.2432G}{\color{blue}465}, 2432

\bibitem[{{Hawcroft} {et~al.}(2024){Hawcroft}, {Sana}, {Mahy}, {Sundqvist}, {de Koter}, {Crowther}, {Bestenlehner}, {Brands}, {David-Uraz}, {Decin}, {Erba}, {Garcia}, {Hamann}, {Herrero}, {Ignace}, {Kee}, {Kub{\'a}tov{\'a}}, {Lefever}, {Moffat}, {Najarro}, {Oskinova}, {Pauli}, {Prinja}, {Puls}, {Sander}, {Shenar}, {St-Louis}, {ud-Doula}, \& {Vink}}]{2024A&A...688A.105H}
{Hawcroft}, C., {Sana}, H., {Mahy}, L., {et~al.} 2024, \href{http://doi.org/10.1051/0004-6361/202245588}{\color{blue}\aap}, \href{https://ui.adsabs.harvard.edu/abs/2024A&A...688A.105H}{\color{blue}688}, A105

\bibitem[{{Jermyn} \& {Cantiello}(2021)}]{2021ApJ...923..104J}
{Jermyn}, A.~S., \& {Cantiello}, M. 2021, \href{http://doi.org/10.3847/1538-4357/ac2d2a}{\color{blue}\apj}, \href{https://ui.adsabs.harvard.edu/abs/2021ApJ...923..104J}{\color{blue}923}, 104

\bibitem[{{Keszthelyi} {et~al.}(2019){Keszthelyi}, {Meynet}, {Georgy}, {Wade}, {Petit}, \& {David-Uraz}}]{2019MNRAS.485.5843K}
{Keszthelyi}, Z., {Meynet}, G., {Georgy}, C., {et~al.} 2019, \href{http://doi.org/10.1093/mnras/stz772}{\color{blue}\mnras}, \href{https://ui.adsabs.harvard.edu/abs/2019MNRAS.485.5843K}{\color{blue}485}, 5843

\bibitem[{{Keszthelyi} {et~al.}(2020){Keszthelyi}, {Meynet}, {Shultz}, {David-Uraz}, {ud-Doula}, {Townsend}, {Wade}, {Georgy}, {Petit}, \& {Owocki}}]{2020MNRAS.493..518K}
{Keszthelyi}, Z., {Meynet}, G., {Shultz}, M.~E., {et~al.} 2020, \href{http://doi.org/10.1093/mnras/staa237}{\color{blue}\mnras}, \href{https://ui.adsabs.harvard.edu/abs/2020MNRAS.493..518K}{\color{blue}493}, 518

\bibitem[{{Keszthelyi} {et~al.}(2022){Keszthelyi}, {de Koter}, {G{\"o}tberg}, {Meynet}, {Brands}, {Petit}, {Carrington}, {David-Uraz}, {Geen}, {Georgy}, {Hirschi}, {Puls}, {Ramalatswa}, {Shultz}, \& {ud-Doula}}]{2022MNRAS.517.2028K}
{Keszthelyi}, Z., {de Koter}, A., {G{\"o}tberg}, Y., {et~al.} 2022, \href{http://doi.org/10.1093/mnras/stac2598}{\color{blue}\mnras}, \href{https://ui.adsabs.harvard.edu/abs/2022MNRAS.517.2028K}{\color{blue}517}, 2028

\bibitem[{{Lecoanet} {et~al.}(2022){Lecoanet}, {Bowman}, \& {Van Reeth}}]{2022MNRAS.512L..16L}
{Lecoanet}, D., {Bowman}, D.~M., \& {Van Reeth}, T. 2022, \href{http://doi.org/10.1093/mnrasl/slac013}{\color{blue}\mnras}, \href{https://ui.adsabs.harvard.edu/abs/2022MNRAS.512L..16L}{\color{blue}512}, L16

\bibitem[{{Marcolino} {et~al.}(2024){Marcolino}, {Bouret}, {Martins}, \& {Hillier}}]{2024A&A...690A.318M}
{Marcolino}, W., {Bouret}, J.~C., {Martins}, F., {et~al.} 2024, \href{http://doi.org/10.1051/0004-6361/202451540}{\color{blue}\aap}, \href{https://ui.adsabs.harvard.edu/abs/2024A&A...690A.318M}{\color{blue}690}, A318

\bibitem[{{Markova} \& {Puls}(2008)}]{2008A&A...478..823M}
{Markova}, N., \& {Puls}, J. 2008, \href{http://doi.org/10.1051/0004-6361:20077919}{\color{blue}\aap}, \href{https://ui.adsabs.harvard.edu/abs/2008A&A...478..823M}{\color{blue}478}, 823

\bibitem[{{Martin} {et~al.}(2018){Martin}, {Neiner}, {Oksala}, {Wade}, {Keszthelyi}, {Fossati}, {Marcolino}, {Mathis}, \& {Georgy}}]{2018MNRAS.475.1521M}
{Martin}, A.~J., {Neiner}, C., {Oksala}, M.~E., {et~al.} 2018, \href{http://doi.org/10.1093/mnras/stx3264}{\color{blue}\mnras}, \href{https://ui.adsabs.harvard.edu/abs/2018MNRAS.475.1521M}{\color{blue}475}, 1521

\bibitem[{{Martins} {et~al.}(2024){Martins}, {Bouret}, {Hillier}, {Brands}, {Crowther}, {Herrero}, {Najarro}, {Pauli}, {Puls}, {Ramachandran}, {Sander}, {Vink}, \& {XShootU Collaboration}}]{2024A&A...689A..31M}
{Martins}, F., {Bouret}, J.~C., {Hillier}, D.~J., {et~al.} 2024, \href{http://doi.org/10.1051/0004-6361/202449457}{\color{blue}\aap}, \href{https://ui.adsabs.harvard.edu/abs/2024A&A...689A..31M}{\color{blue}689}, A31

\bibitem[{{Naz{\'e}} {et~al.}(2007){Naz{\'e}}, {Rauw}, {Pollock}, {Walborn}, \& {Howarth}}]{2007MNRAS.375..145N}
{Naz{\'e}}, Y., {Rauw}, G., {Pollock}, A.~M.~T., {et~al.} 2007, \href{http://doi.org/10.1111/j.1365-2966.2006.11270.x}{\color{blue}\mnras}, \href{https://ui.adsabs.harvard.edu/abs/2007MNRAS.375..145N}{\color{blue}375}, 145

\bibitem[{{Oksala} {et~al.}(2012){Oksala}, {Wade}, {Townsend}, {Owocki}, {Kochukhov}, {Neiner}, {Alecian}, \& {Grunhut}}]{2012MNRAS.419..959O}
{Oksala}, M.~E., {Wade}, G.~A., {Townsend}, R.~H.~D., {et~al.} 2012, \href{http://doi.org/10.1111/j.1365-2966.2011.19753.x}{\color{blue}\mnras}, \href{https://ui.adsabs.harvard.edu/abs/2012MNRAS.419..959O}{\color{blue}419}, 959

\bibitem[{{Parsons} {et~al.}(2024){Parsons}, {Prinja}, {Bernini-Peron}, {Fullerton}, {Massa}, {Oskinova}, {Pauli}, {Rickard}, \& {Sander}}]{2024MNRAS.52711422P}
{Parsons}, T.~N., {Prinja}, R.~K., {Bernini-Peron}, M., {et~al.} 2024, \href{http://doi.org/10.1093/mnras/stad3966}{\color{blue}\mnras}, \href{https://ui.adsabs.harvard.edu/abs/2024MNRAS.52711422P}{\color{blue}527}, 11422

\bibitem[{{Petit} \& {Wade}(2012)}]{2012MNRAS.420..773P}
{Petit}, V., \& {Wade}, G.~A. 2012, \href{http://doi.org/10.1111/j.1365-2966.2011.20091.x}{\color{blue}\mnras}, \href{https://ui.adsabs.harvard.edu/abs/2012MNRAS.420..773P}{\color{blue}420}, 773

\bibitem[{{Petit} {et~al.}(2017){Petit}, {Keszthelyi}, {MacInnis}, {Cohen}, {Townsend}, {Wade}, {Thomas}, {Owocki}, {Puls}, \& {ud-Doula}}]{2017MNRAS.466.1052P}
{Petit}, V., {Keszthelyi}, Z., {MacInnis}, R., {et~al.} 2017, \href{http://doi.org/10.1093/mnras/stw3126}{\color{blue}\mnras}, \href{https://ui.adsabs.harvard.edu/abs/2017MNRAS.466.1052P}{\color{blue}466}, 1052

\bibitem[{{Schneider} {et~al.}(2020){Schneider}, {Ohlmann}, {Podsiadlowski}, {R{\"o}pke}, {Balbus}, \& {Pakmor}}]{2020MNRAS.495.2796S}
{Schneider}, F.~R.~N., {Ohlmann}, S.~T., {Podsiadlowski}, P., {et~al.} 2020, \href{http://doi.org/10.1093/mnras/staa1326}{\color{blue}\mnras}, \href{https://ui.adsabs.harvard.edu/abs/2020MNRAS.495.2796S}{\color{blue}495}, 2796

\bibitem[{{Schneider} {et~al.}(2019){Schneider}, {Ohlmann}, {Podsiadlowski}, {R{\"o}pke}, {Balbus}, {Pakmor}, \& {Springel}}]{2019Natur.574..211S}
{Schneider}, F. R.~N., {Ohlmann}, S.~T., {Podsiadlowski}, P., {et~al.} 2019, \href{http://doi.org/10.1038/s41586-019-1621-5}{\color{blue}\nat}, \href{https://ui.adsabs.harvard.edu/abs/2019Natur.574..211S}{\color{blue}574}, 211

\bibitem[{{Schuessler} \& {Paehler}(1978)}]{1978A&A....68...57S}
{Schuessler}, M., \& {Paehler}, A. 1978, \aap, \href{https://ui.adsabs.harvard.edu/abs/1978A&A....68...57S}{\color{blue}68}, 57

\bibitem[{{Shenar} {et~al.}(2023){Shenar}, {Wade}, {Marchant}, {Bagnulo}, {Bodensteiner}, {Bowman}, {Gilkis}, {Langer}, {Nicolas-Chen{\'e}}, {Oskinova}, {Van Reeth}, {Sana}, {St-Louis}, {de Oliveira}, {Todt}, \& {Toonen}}]{2023Sci...381..761S}
{Shenar}, T., {Wade}, G.~A., {Marchant}, P., {et~al.} 2023, \href{http://doi.org/10.1126/science.ade3293}{\color{blue}Science}, \href{https://ui.adsabs.harvard.edu/abs/2023Sci...381..761S}{\color{blue}381}, 761

\bibitem[{{Shultz} {et~al.}(2015){Shultz}, {Wade}, {Alecian}, \& {BinaMIcS Collaboration}}]{2015MNRAS.454L...1S}
{Shultz}, M., {Wade}, G.~A., {Alecian}, E., {et~al.} 2015, \href{http://doi.org/10.1093/mnrasl/slv096}{\color{blue}\mnras}, \href{https://ui.adsabs.harvard.edu/abs/2015MNRAS.454L...1S}{\color{blue}454}, L1

\bibitem[{{Shultz} {et~al.}(2019{\natexlab{a}}){Shultz}, {Le Bouquin}, {Rivinius}, {Wade}, {Kochukhov}, {Alecian}, {Petit}, {Pfuhl}, {Karl}, {Gao}, {Grellmann}, {Lin}, {Garcia}, {Lacour}, {MiMeS Collaboration}, \& {BinaMIcS Collaboration}}]{2019MNRAS.482.3950S}
{Shultz}, M., {Le Bouquin}, J.~B., {Rivinius}, T., {et~al.} 2019{\natexlab{a}}, \href{http://doi.org/10.1093/mnras/sty2985}{\color{blue}\mnras}, \href{https://ui.adsabs.harvard.edu/abs/2019MNRAS.482.3950S}{\color{blue}482}, 3950

\bibitem[{{Shultz} {et~al.}(2019{\natexlab{b}}){Shultz}, {Wade}, {Rivinius}, {Alecian}, {Neiner}, {Petit}, {Owocki}, {ud-Doula}, {Kochukhov}, {Bohlender}, {Keszthelyi}, {MiMeS Collaboration}, \& {BinaMIcS Collaboration}}]{2019MNRAS.490..274S}
{Shultz}, M.~E., {Wade}, G.~A., {Rivinius}, T., {et~al.} 2019{\natexlab{b}}, \href{http://doi.org/10.1093/mnras/stz2551}{\color{blue}\mnras}, \href{https://ui.adsabs.harvard.edu/abs/2019MNRAS.490..274S}{\color{blue}490}, 274

\bibitem[{{Sikora} {et~al.}(2019){Sikora}, {Wade}, {Power}, \& {Neiner}}]{2019MNRAS.483.3127S}
{Sikora}, J., {Wade}, G.~A., {Power}, J., {et~al.} 2019, \href{http://doi.org/10.1093/mnras/sty2895}{\color{blue}\mnras}, \href{https://ui.adsabs.harvard.edu/abs/2019MNRAS.483.3127S}{\color{blue}483}, 3127

\bibitem[{{Silvester} {et~al.}(2014){Silvester}, {Kochukhov}, \& {Wade}}]{2014MNRAS.440..182S}
{Silvester}, J., {Kochukhov}, O., \& {Wade}, G.~A. 2014, \href{http://doi.org/10.1093/mnras/stu306}{\color{blue}\mnras}, \href{https://ui.adsabs.harvard.edu/abs/2014MNRAS.440..182S}{\color{blue}440}, 182

\bibitem[{{Stibbs}(1950)}]{1950MNRAS.110..395S}
{Stibbs}, D.~W.~N. 1950, \href{http://doi.org/10.1093/mnras/110.4.395}{\color{blue}\mnras}, \href{https://ui.adsabs.harvard.edu/abs/1950MNRAS.110..395S}{\color{blue}110}, 395

\bibitem[{{Sundqvist} {et~al.}(2013){Sundqvist}, {Petit}, {Owocki}, {Wade}, {Puls}, \& {MiMeS Collaboration}}]{2013MNRAS.433.2497S}
{Sundqvist}, J.~O., {Petit}, V., {Owocki}, S.~P., {et~al.} 2013, \href{http://doi.org/10.1093/mnras/stt921}{\color{blue}\mnras}, \href{https://ui.adsabs.harvard.edu/abs/2013MNRAS.433.2497S}{\color{blue}433}, 2497

\bibitem[{{ud-Doula} \& {Owocki}(2002)}]{2002ApJ...576..413U}
{ud-Doula}, A., \& {Owocki}, S.~P. 2002, \href{http://doi.org/10.1086/341543}{\color{blue}\apj}, \href{https://ui.adsabs.harvard.edu/abs/2002ApJ...576..413U}{\color{blue}576}, 413

\bibitem[{{ud-Doula} {et~al.}(2009){ud-Doula}, {Owocki}, \& {Townsend}}]{2009MNRAS.392.1022U}
{ud-Doula}, A., {Owocki}, S.~P., \& {Townsend}, R. H.~D. 2009, \href{http://doi.org/10.1111/j.1365-2966.2008.14134.x}{\color{blue}\mnras}, \href{https://ui.adsabs.harvard.edu/abs/2009MNRAS.392.1022U}{\color{blue}392}, 1022

\bibitem[{{Villebrun} {et~al.}(2019){Villebrun}, {Alecian}, {Hussain}, {Bouvier}, {Folsom}, {Lebreton}, {Amard}, {Charbonnel}, {Gallet}, {Haemmerl{\'e}}, {B{\"o}hm}, {Johns-Krull}, {Kochukhov}, {Marsden}, {Morin}, \& {Petit}}]{2019A&A...622A..72V}
{Villebrun}, F., {Alecian}, E., {Hussain}, G., {et~al.} 2019, \href{http://doi.org/10.1051/0004-6361/201833545}{\color{blue}\aap}, \href{https://ui.adsabs.harvard.edu/abs/2019A&A...622A..72V}{\color{blue}622}, A72

\bibitem[{{Wade} {et~al.}(2016){Wade}, {Neiner}, {Alecian}, {Grunhut}, {Petit}, {Batz}, {Bohlender}, {Cohen}, {Henrichs}, {Kochukhov}, {Landstreet}, {Manset}, {Martins}, {Mathis}, {Oksala}, {Owocki}, {Rivinius}, {Shultz}, {Sundqvist}, {Townsend}, {ud-Doula}, {Bouret}, {Braithwaite}, {Briquet}, {Carciofi}, {David-Uraz}, {Folsom}, {Fullerton}, {Leroy}, {Marcolino}, {Moffat}, {Naz{\'e}}, {Louis}, {Auri{\`e}re}, {Bagnulo}, {Bailey}, {Barb{\'a}}, {Blaz{\`e}re}, {B{\"o}hm}, {Catala}, {Donati}, {Ferrario}, {Harrington}, {Howarth}, {Ignace}, {Kaper}, {L{\"u}ftinger}, {Prinja}, {Vink}, {Weiss}, \& {Yakunin}}]{2016MNRAS.456....2W}
{Wade}, G.~A., {Neiner}, C., {Alecian}, E., {et~al.} 2016, \href{http://doi.org/10.1093/mnras/stv2568}{\color{blue}\mnras}, \href{https://ui.adsabs.harvard.edu/abs/2016MNRAS.456....2W}{\color{blue}456}, 2

\end{thebibliography}

\end{document}